\documentclass[preprintnumbers,10pt,nofootinbib]{revtex4}

\usepackage{amsmath,latexsym,amssymb,amsfonts}
\usepackage[dvips]{graphicx,color}
\usepackage{bm}

\begin{document}

\preprint{CTPU-PTC-18-25}

\title{\textbf{Quantum creation of traversable wormholes \textit{ex nihilo}\\ in Gauss-Bonnet-dilaton gravity}}
\author{
\textsc{Gansukh Tumurtushaa$^{a}$}\footnote{{\tt gansuhmgl@ibs.re.kr}}
and
\textsc{Dong-han Yeom$^{b,c,d,e}$}\footnote{{\tt innocent.yeom{}@{}gmail.com}}
}

\affiliation{
$^{a}$\small{Center for Theoretical Physics of the Universe, Institute for Basic Science, Daejeon 34051, Republic of Korea}\\
$^{b}$\small{Asia Pacific Center for Theoretical Physics, Pohang 37673, Republic of Korea}\\
$^{c}$\small{Department of Physics, POSTECH, Pohang 37673, Republic of Korea}\\
$^{d}$\small{Department of Physics Education, Pusan National University, Busan 46241, Republic of Korea}\\
$^{e}$\small{Research Center for Dielectric and Advanced Matter Physics, Pusan National University, Busan 46241, Republic of Korea}
}

\begin{abstract}
We investigate a nucleation of a Euclidean wormhole and its analytic continuation to Lorentzian signatures in Gauss-Bonnet-dilaton gravity, where this model can be embedded by the type-II superstring theory. We show that there exists a Euclidean wormhole solution in this model by choosing a suitable shape of the dilaton potential. After the analytic continuation, this explains a quantum creation of a time-like traversable wormhole. Finally, we discuss relations to the information loss problem and the current literature.
\end{abstract}

\maketitle

\newpage

\tableofcontents


\section{Introduction}

The investigation of wormholes is a very hypothetical but interesting research topic. One may study wormholes in the Lorentzian signatures; if the wormhole throat is time-like and traversable, then it should violate the averaged null energy condition by a certain way, e.g., introducing an exotic matter or quantum effects \cite{Morris:1988cz}. One may also study wormholes in the Euclidean signatures \cite{Hawking:1988ae}; then this topic can have applications for quantum gravity \cite{Coleman:1988tj} and quantum cosmology \cite{Chen:2016ask}.

For both cases, Lorentzian or Euclidean wormholes, one needs to introduce an exotic matter in order to obtain a wormhole throat. However, we have two troubles with this (for more details, see \cite{Visser:1995cc}).
\begin{itemize}
\item[(1)] Can we introduce the (at least, effective) exotic matter within the well-known safe field theory without the menace of phantom \cite{Cline:2003gs}?
\item[(2)] Why should the two separated spaces be joined at the throat, i.e., is there a mechanism to bring two spaces to one place?
\end{itemize}
These two are challenging but difficult questions.

However, recently, especially by string theorists, it was suggested that wormholes will do an important role within the string theory. For example, in order to overcome the inconsistency \cite{Yeom:2008qw,Almheiri:2012rt} of black hole complementarity \cite{Susskind:1993if} and resolve the firewall paradox \cite{Hwang:2012nn}, the Einstein-Rosen/Einstein-Podolsky-Rosen (ER=EPR) conjecture has been suggested \cite{Maldacena:2013xja}, where this is related to a space-like wormhole with the Einstein-Rosen bridge. Later, it was known that semi-classical effects on the Einstein-Rosen bridge can change the bridge to be traversable \cite{Chen:2016nvj}. Moreover, recently, it was reported that the existence of wormholes can be allowed within the string theory as we turn on suitable semi-classical quantum effects \cite{Gao:2016bin}.

Therefore, it is interesting to study whether there can be other time-like wormhole solutions within the string theory or not. In this paper, we will focus on the Gauss-Bonnet-dilaton gravity, where this can be embedded by the string theory \cite{Metsaev:1987zx}. Previously, it was reported that there exists Lorentzian wormhole solutions within the theory \cite{Kanti:2011jz}. This is quite impressive since now we can say that the string theory seems to allow time-like wormholes even at the classical level. Also, energy conditions will be effectively violated, but we may trust the model since the string theory can be regarded as a candidate of the UV-completion of gravity.

On the other hand, for all models of wormholes, their dynamical constructions were not known yet. In this paper, we will argue that Euclidean wormholes can be analytically continued to a time-like Lorentzian wormhole. We will find a Euclidean wormhole solution in the Gauss-Bonnet-dilaton gravity, and hence, the solution will be allowed by the string theory context. Therefore, based on our model, we can solve two previous difficult questions: (1) we used a string-inspired model which can be regarded as a UV-completion and (2) we explain a dynamical creation of a wormhole $\textit{ex nihilo}$ by quantum mechanical processes.

This paper is constructed as follows. In Sec.~\ref{sec:mod}, we describe the details of the Gauss-Bonnet-dilaton gravity model. In Sec.~\ref{sec:sol}, we describe the details of the solution as well as its properties. In Sec.~\ref{sec:dis}, we give some comments on the applications of wormholes and discuss possible future works.

\section{\label{sec:mod}Model}

The four-dimensional string effective action can be expanded as follows \cite{Metsaev:1987zx}:
\begin{eqnarray}
S = \int d^{4}x \sqrt{-g} \left[ \frac{R}{2\kappa^{2}} - \frac{1}{2} \left(\nabla \phi\right)^{2} - V(\phi) + \frac{\lambda}{2} e^{-c \phi} \left( \mathcal{C}_{1} R_{\mathrm{GB}}^{2} + \mathcal{C}_{2} \left( \nabla \phi \right)^{4} \right) \right],
\end{eqnarray}
where $R$ is the Ricci scalar, $\kappa^{2} = 8\pi$, $\phi$ is the dilaton field with potential $V(\phi)$, $c$ is a constant, $\lambda$ is the coupling constant proportional to the $\alpha'$ parameter, and
\begin{eqnarray}
R_{\mathrm{GB}}^{2} = R_{\mu\nu\rho\sigma}R^{\mu\nu\rho\sigma} - 4 R_{\mu\nu}R^{\mu\nu} + R^{2}
\end{eqnarray}
is the Gauss-Bonnet term. Here, $\mathcal{C}_{1}$ and $\mathcal{C}_{2}$ are model dependent parameters, where its ratio $\mathcal{C}_{2}/\mathcal{C}_{1}$ is $2:0:1$ for three types of string theories, i.e., the bosonic string theory, the type-II superstring theory, and the heterotic superstring theory, respectively \cite{Kanti:1995vq}.

Therefore, the type-II superstring theory includes the Gauss-Bonnet-dilaton gravity model:
\begin{eqnarray}
S = \int d^{4}x \sqrt{-g} \left[ \frac{R}{2\kappa^{2}} - \frac{1}{2} \left(\nabla \phi\right)^{2} - V(\phi) + \frac{1}{2} \xi(\phi) R_{\mathrm{GB}}^{2} \right],
\end{eqnarray}
where
\begin{eqnarray}
\xi(\phi) = \lambda e^{- c (\phi-\phi_{0})}
\end{eqnarray}
with constants $\lambda$, $c$, and $\phi_{0}$ without loss of generality.

In order to investigate non-perturbative effects of the model, we introduce the Hartle-Hawking wave function as a Euclidean path-integral \cite{Hartle:1983ai}:
\begin{eqnarray}
\Psi\left[ h_{\mu\nu}, \psi \right] = \int \mathcal{D}g_{\mu\nu} \mathcal{D} \phi \;\; e^{- S_{\mathrm{E}}\left[g_{\mu\nu}, \phi \right]},
\end{eqnarray}
where we sum-over all regular Euclidean field combinations that satisfy $\partial g_{\mu\nu} = h_{\mu\nu}$ and $\partial \phi = \psi$. By using the steepest-descent approximation, this path-integral will be well approximated by Euclidean on-shell solutions, or so-called instantons \cite{Hartle:1983ai,Hartle:2007gi,Hwang:2011mp}. Especially, we investigate instantons with the following $O(4)$-symmetric metric ansatz:
\begin{eqnarray}
ds_{\mathrm{E}}^{2} = d\tau^{2} + a^{2}(\tau) d\Omega_{3}^{2},
\end{eqnarray}
where
\begin{eqnarray}
d\Omega_{3}^{2} = d\chi^{2} + \sin^{2}{\chi} \left( d\theta^{2} + \sin^{2}{\theta} d\varphi^{2} \right)
\end{eqnarray}
is the three-sphere.

\subsection{Equations of motion}

We can present equations of motion in Euclidean signatures \cite{Koh:2014bka}:
\begin{eqnarray}
\label{Eq1} H^{2} &=& \frac{\kappa^{2}}{3} \left[ \frac{1}{2} \dot{\phi}^{2} - V + \frac{3K}{\kappa^{2} a^{2}} - 12 \dot{\xi} H  \left( - H^{2} + \frac{K}{a^{2}} \right) \right],\\
\label{Eq2} \dot{H} &=& -\frac{\kappa^{2}}{2} \left[ \dot{\phi}^{2} + \frac{2K}{\kappa^{2} a^{2}} + 4\ddot{\xi} \left(-H^{2} + \frac{K}{a^{2}} \right) + 4 \dot{\xi} H \left(-2 \dot{H} + H^{2} - \frac{3K}{a^{2}} \right) \right],\\
\label{Eq3} 0 &=& \ddot{\phi} + 3 H \dot{\phi} - V' - 12 \xi' \left( - H^{2} + \frac{K}{a^{2}} \right) \left( \dot{H} + H^{2} \right),
\end{eqnarray}
where $H = \dot{a}/a$, $K = +1$, and $\kappa^{2} = 8\pi$. This set of equations are consistent with \cite{Ro:2016kyu}.

Eqs.~(\ref{Eq2}) and (\ref{Eq3}) will be used to numerically solve the variables. Eq.~(\ref{Eq1}) will be the constraint equation, where this is simplified (if $a \neq 0$) to
\begin{eqnarray}
0 = 6a \left( K - \dot{a}^{2} \right) - 24 \kappa^{2} \dot{a} \left( K - \dot{a}^{2} \right) \dot{\phi} \xi' + \kappa^{2} a^{3} \left( \dot{\phi}^{2} - 2V \right).
\end{eqnarray}
We present Eqs.~(\ref{Eq2}) and (\ref{Eq3}) for $\ddot{a}$ and $\ddot{\phi}$.
\begin{eqnarray}
\label{eq:a}\ddot{a} &=& - \frac{a^{2}}{2} \mathcal{F},\\
\label{eq:phi}\ddot{\phi} &=& V' - 3 \frac{\dot{a}}{a} \dot{\phi} - \frac{6 \xi' \left(K - \dot{a}^{2}\right)}{a} \mathcal{F},
\end{eqnarray}
where
\begin{equation}
\mathcal{F} \equiv \frac{2a (K-\dot{a}^{2}) + \kappa^{2} a^{3} \dot{\phi}^{2} - 4\kappa^{2}\xi''(- K a \dot{\phi}^{2} + a \dot{a}^{2} \dot{\phi}^{2} ) - 4\kappa^{2} \xi'( - K a V' + a \dot{a}^{2} V' + 6 K \dot{a} \dot{\phi} - 6 \dot{a}^{3} \dot{\phi})}{a^{4} - 4\kappa^{2} a^{3} \dot{a} \dot{\phi} \xi' + 24 K^{2} \kappa^{2} \xi'^{2} - 48 K \kappa^{2} \dot{a}^{2} \xi'^{2} + 24 \kappa^{2} \dot{a}^{4} \xi'^{2}}.
\end{equation}
These expressions are useful since the right-hand sides of Eqs.~(\ref{eq:a}) and (\ref{eq:phi}) are functionals up to first derivations, i.e., $a$, $\dot{a}$, $\phi$, and $\dot{\phi}$.

\subsection{Solving techniques}

Usually, one gives the initial condition at a certain point and solve equations. One important point is to give regular boundary conditions for the end of the solution if the solution is compact. In order to make this boundary value problem simpler, we solve $V$ as a function of $\tau$ rather than directly solve $\phi$. This means that we first fix the solution $\phi$ and solve equations for $a$ and $V$.

\begin{figure}
\begin{center}
\includegraphics[scale=0.75]{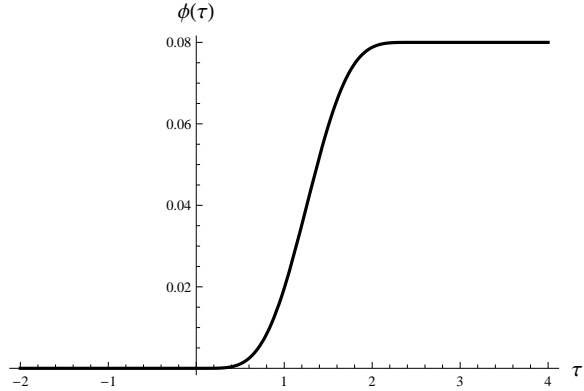}
\caption{\label{fig:phi}An example of $\phi(\tau)$, where $\phi_{0} = 0$, $\phi_{1} = 0.08$, and $\Delta = 0.8$.}
\end{center}
\end{figure}

As we solve $V$ instead of $\phi$, one can arbitrarily choose the shape of the field in principle and we use
\begin{eqnarray}
\phi(\tau) = \phi_{0} + \frac{\left(\phi_{1} - \phi_{0} \right)}{12\pi} \left[ 12 \frac{\phi - \phi_{0}}{\Delta} - 8\sin 2\frac{\phi - \phi_{0}}{\Delta} + \sin 4 \frac{\phi - \phi_{0}}{\Delta} \right]
\end{eqnarray}
for $0 \leq \phi - \phi_{0} \leq \pi\Delta$, while $\phi(\tau) = \phi_{1}$ for $\phi-\phi_{0} > \pi\Delta$ and $\phi(\tau) = \phi_{0}$ for $\phi-\phi_{0} < \pi\Delta$ (Fig.~\ref{fig:phi}). This function is continuous up to third order differentiations. Here, $\phi_{0}$, $\phi_{1}$, and $\Delta$ are free parameters.

Finally, the equation for $V$ becomes $\dot{V} = \dot{\phi} V'$, where
\begin{equation}
V' = \frac{a^{5} \ddot{\phi} + 6 \kappa^{2} a^{3} (K - 3 \dot{a}^{2}) \dot{\phi}^{2} \xi' - 12 \xi' (K - \dot{a}^{2})^{2} ( 6 \kappa^{2} \dot{a} \dot{\phi} \xi' - a - 2 a \kappa^{2} \dot{\phi}^{2} \xi'' - 2 a \kappa^{2} \ddot{\phi} \xi') + a^{4} \dot{a} \dot{\phi} (3 - 4 \kappa^{2} \ddot{\phi} \xi')}{a^{4} (a - 4\kappa^{2} \dot{a} \dot{\phi} \xi')}.
\end{equation}

In the first glimpse, this is not due to the legitimate choice of the potential shape; rather, the potential is chosen by the shape of the solution. However, this is still a good strategy (for example, see \cite{Kanno:2012zf}) if (1) we do not know the exact shape of the potential of the theory from the first principle and (2) we would like to know whether a specific type of solutions does exist or not. At once one can construct a solution with an arbitrary shape of the potential, then the theory does not disallow such a solution in principle; and then the next task is the justification of the potential that allows such a solution. In this paper, we will mainly focus on the first step of the question: the existence of the solution in the Gauss-Bonnet-dilaton gravity.

\subsection{Conditions for wormholes}

In order to have a Euclidean wormhole, we need $\ddot{a} > 0$ when $\dot{a} = 0$ and $a > 0$. The necessary conditions are
\begin{eqnarray}
\ddot{a} &=& - \frac{a^{3}}{2} \left( \frac{2 K + \kappa^{2} \dot{\phi}^{2} (a^{2} + 4 \xi'' K) + 4 \kappa^{2} \xi' K V'}{a^{4} + 24 K^{2} \kappa^{2} \xi'^{2}} \right),\\
\dot{\phi}^{2} &=& - \frac{6K}{\kappa^{2} a^{2}} + 2 V.
\end{eqnarray}
These are equivalent to
\begin{eqnarray}
\frac{1}{2\kappa^{2}} + \xi' V' + \left( \frac{a^{2}}{4K} + \xi'' \right) \dot{\phi}^{2} &<& 0,\\
V - \frac{3 K}{\kappa^{2} a^{2}} &\geq& 0.
\end{eqnarray}
These conditions are useful to figure out the properties of wormholes. This means that the Euclidean wormhole throat can exist only if $V > 0$ and $\xi' V' < 0$ are satisfied.

\begin{figure}
\begin{center}
\includegraphics[scale=0.75]{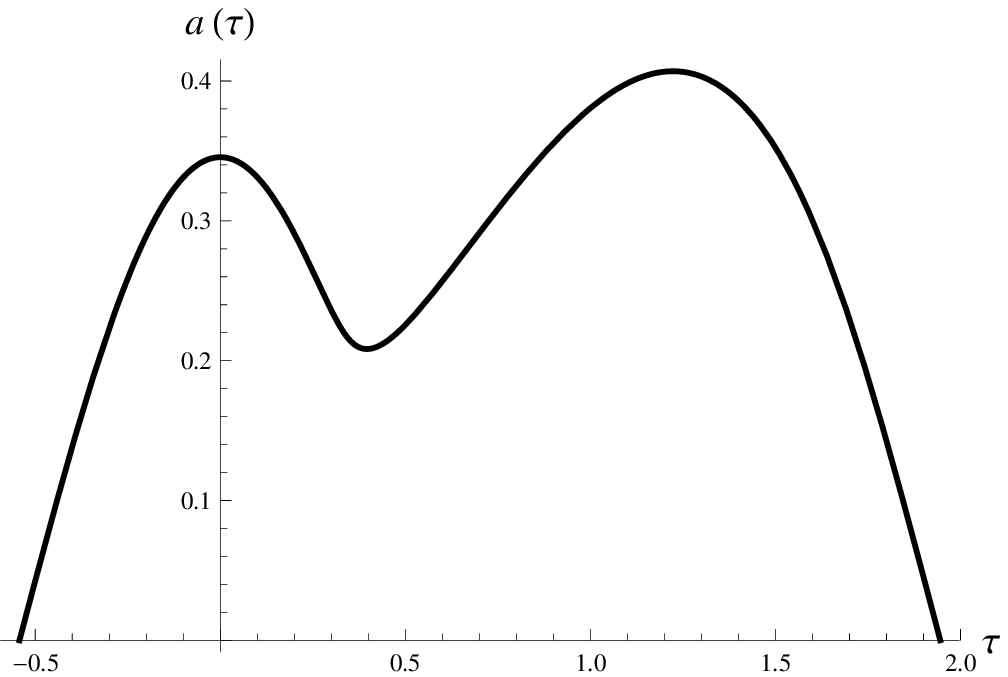}
\includegraphics[scale=0.75]{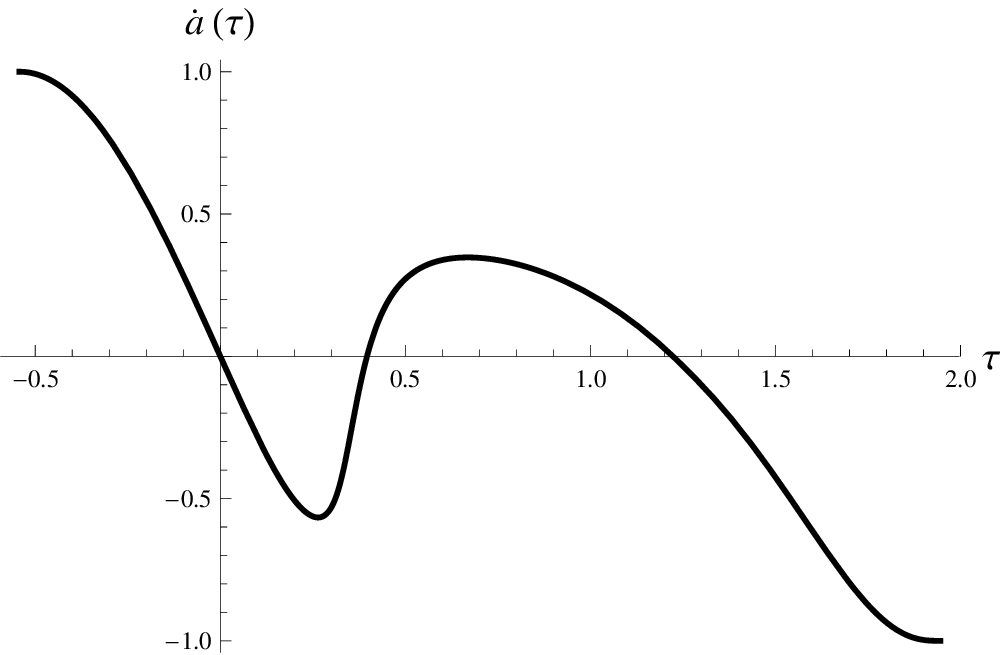}
\caption{\label{fig:atau}$a(\tau)$ (left) and $\dot{a}(\tau)$ (right).}
\includegraphics[scale=0.75]{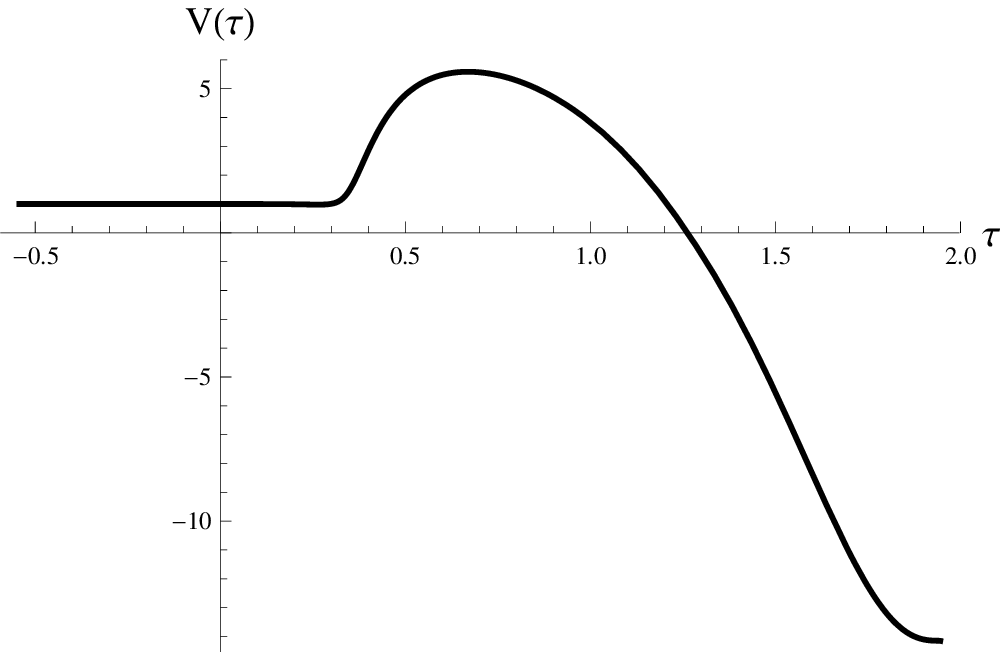}
\caption{\label{fig:Vtau}$V(\tau)$.}
\end{center}
\end{figure}

\subsection{Initial conditions}

For the numerical solving procedure, except the model parameters that we have mentioned, we need to choose the initial conditions:
\begin{eqnarray}
a(0) &=& \sqrt{\frac{6K}{\kappa^{2} \left(2 V_{0} - \dot{\phi}^{2}(0) \right)}},\\
\dot{a}(0) &=& 0,\\
V(0) &=& V_{0},
\end{eqnarray}
where, for convenience, we choose
\begin{eqnarray}
\phi(0) &=& \phi_{0},\\
\dot{\phi}(0) &=& 0.
\end{eqnarray}
Hence, $\tau < 0$ region is a pure de Sitter space with $V = V_{0}$ and $\tau = 0$ is the half-way point ($\dot{a} = 0$ and $\ddot{a} < 0$) of the de Sitter space. $\phi_{0}$ can be chosen to be zero without loss of generality. From this initial condition, we can check whether the Euclidean wormhole solution can appear at $\tau > 0$ or not.

\section{\label{sec:sol}Solutions and properties}

By choosing suitable initial conditions and model parameters, one can obtain a Euclidean wormhole with regular boundary conditions at compact boundaries $a = 0$. We demonstrate an explicit example.

\subsection{Quantum creation of time-like wormholes}

\begin{figure}
\begin{center}
\includegraphics[scale=0.75]{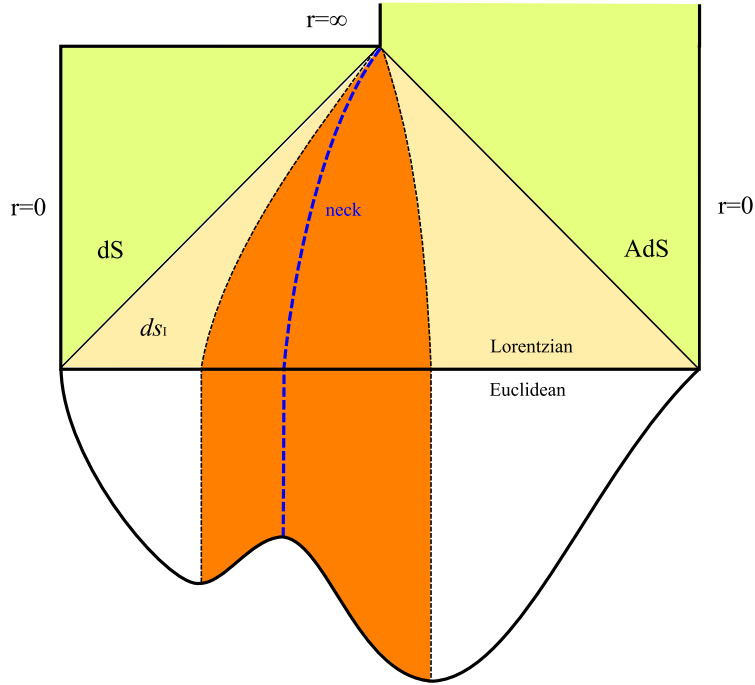}
\caption{\label{fig:concept}The analytic continuation to Lorentzian signatures. Here, $r$ is the areal radius. The triangular part is the region where the metric $ds_{I}$ is applied. In the orange colored part, the spacetime includes a time-like wormhole. The green colored parts satisfy the metric $ds_{II}$, where left one is de Sitter and the right one is anti-de Sitter.}
\end{center}
\end{figure}

We can obtain a Euclidean wormhole with smooth boundary conditions, i.e., $\dot{a} = \pm 1$ for $a = 0$, as in Figs.~\ref{fig:atau} and \ref{fig:Vtau}, where we have chosen $V_{0} = 1$. The left end, $\tau < 0$, of the solution is de Sitter and the right end of the solution, $\tau \sim 2$, is anti-de Sitter. Two compact regions are connected by a Euclidean wormhole, i.e., there exists a $\ddot{a} > 0$ and $\dot{a} = 0$ point. 

After the inhomogeneous Wick-rotation $d\chi = i dT$, one can obtain a time-like wormhole in the Lorentzian signatures \cite{Hawking:1998bn}. If we consider the analytic continuation $\chi = \pi/2 + iT$, then the Lorentzian metric becomes
\begin{eqnarray}\label{eq:metric1}
ds_{I}^{2} = d\tau^{2} + a^{2}(\tau) \left( - dT^{2} + \cosh^{2} T \; d\Omega_{2}^{2} \right).
\end{eqnarray}
In these coordinates, $0 \leq T < \infty$ is the time-like parameter (constant $T$ surfaces are space-like) and $\tau$ is the space-like parameter (constant $\tau$ surfaces are time-like). Note that both ends of $a = 0$ correspond null surfaces. Beyond the null surfaces, one can further analytically continue by choosing $\tau = it$, $T = i\pi/2 +\chi$, and $\alpha(t) = -ia(it)$:
\begin{eqnarray}
ds_{II}^{2} = - dt^{2} + \alpha^{2}(t) \left( d\chi^{2} + \sinh^{2} \chi \; d\Omega_{2}^{2} \right).
\end{eqnarray}

The analytically continued causal structures are summarized by Fig.~\ref{fig:concept}. The bottle-neck of the wormhole is located at the metric $ds_{I}$, where this corresponds $\dot{a} = 0$ and $\ddot{a} > 0$. For a constant space-like hypersurface $T = \mathrm{const.}$, the areal radius is proportional to $a(\tau)$, and hence, the point with $\dot{a} = 0$ and $\ddot{a} > 0$ satisfies the flare-out condition, where this is the correct condition for the bottle-neck of the wormhole. In addition, $\tau = \mathrm{const.}$ is manifestly time-like and hence the bottle-neck is traversable. One additional comment is that this bottle-neck does not connect two asymptotic infinities; rather, this connects between two maximums of the areal radius (of de Sitter background). 

\begin{figure}
\begin{center}
\includegraphics[scale=0.6]{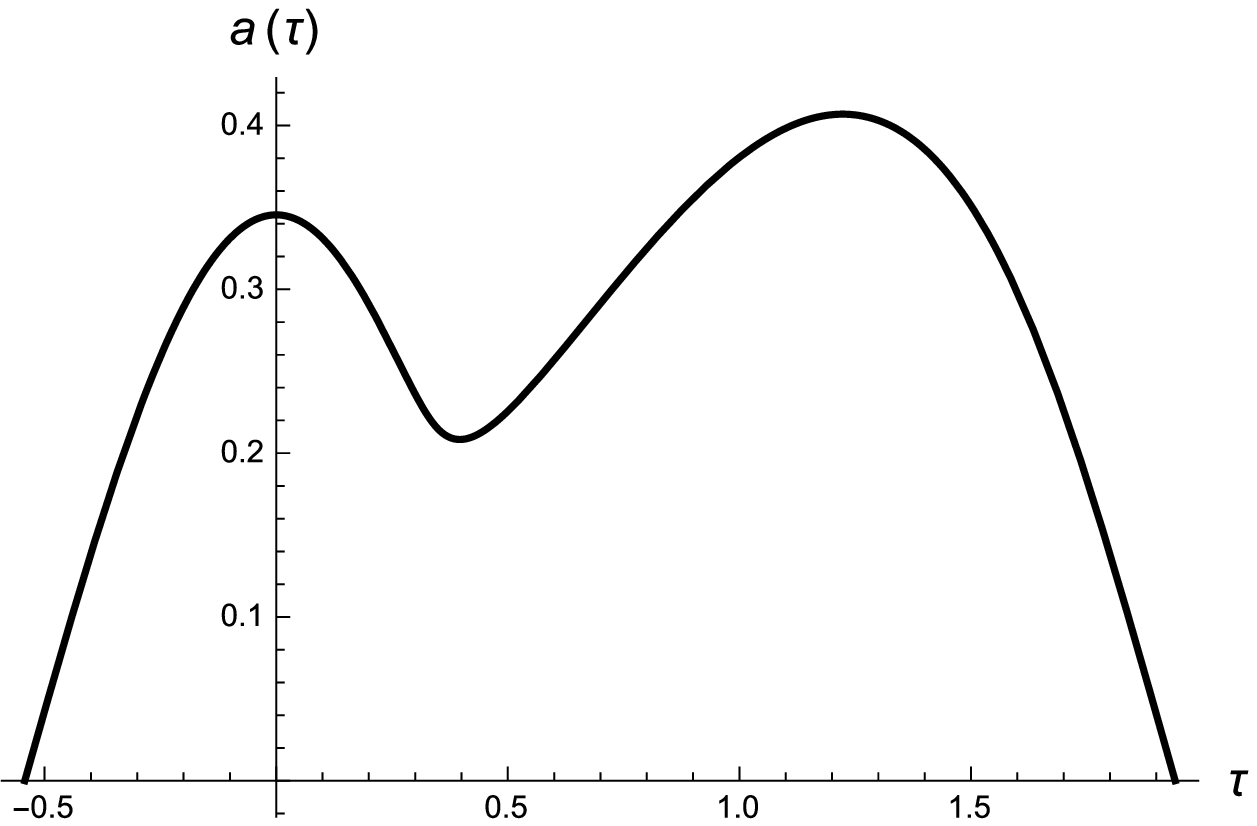}
\includegraphics[scale=0.6]{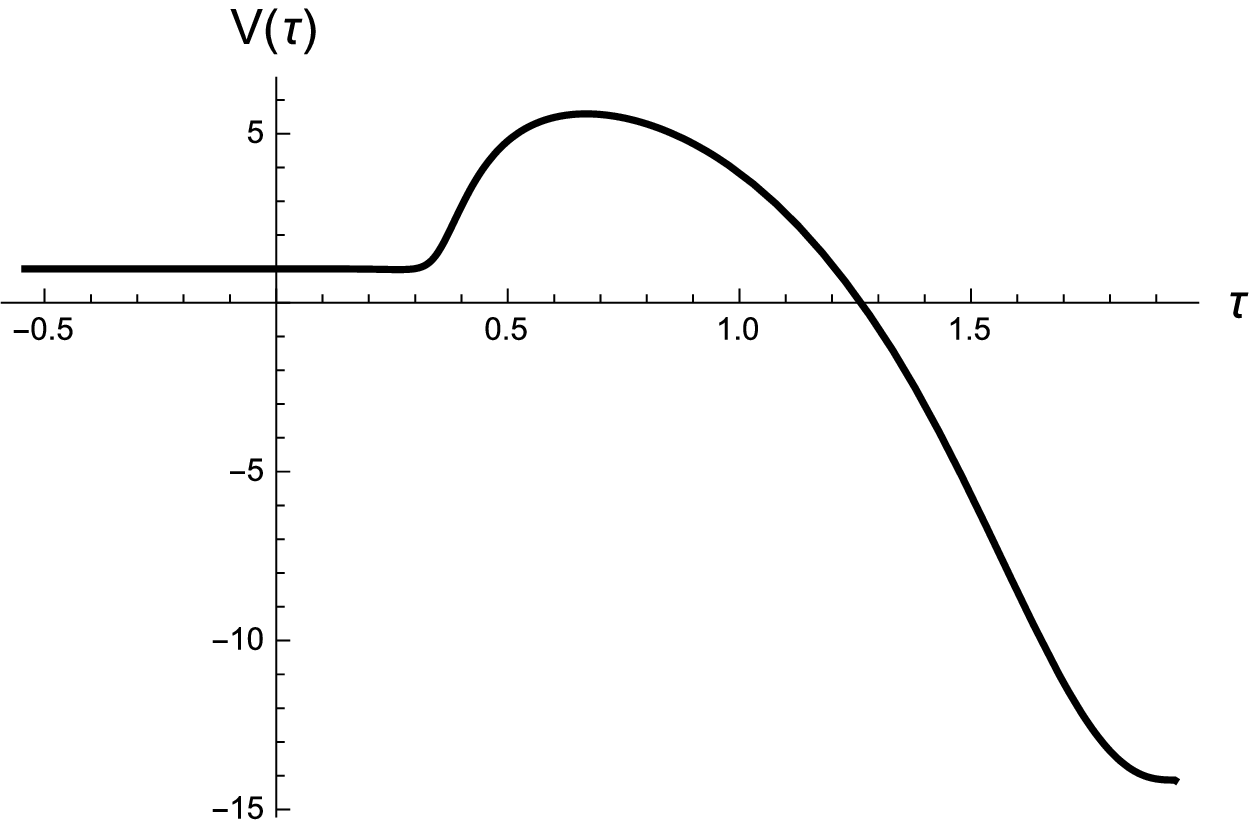}
\caption{\label{fig:demonstrate1}$a(\tau)$ (left) and $V(\tau)$ (right) for $\Delta = 0.8$.}
\includegraphics[scale=0.6]{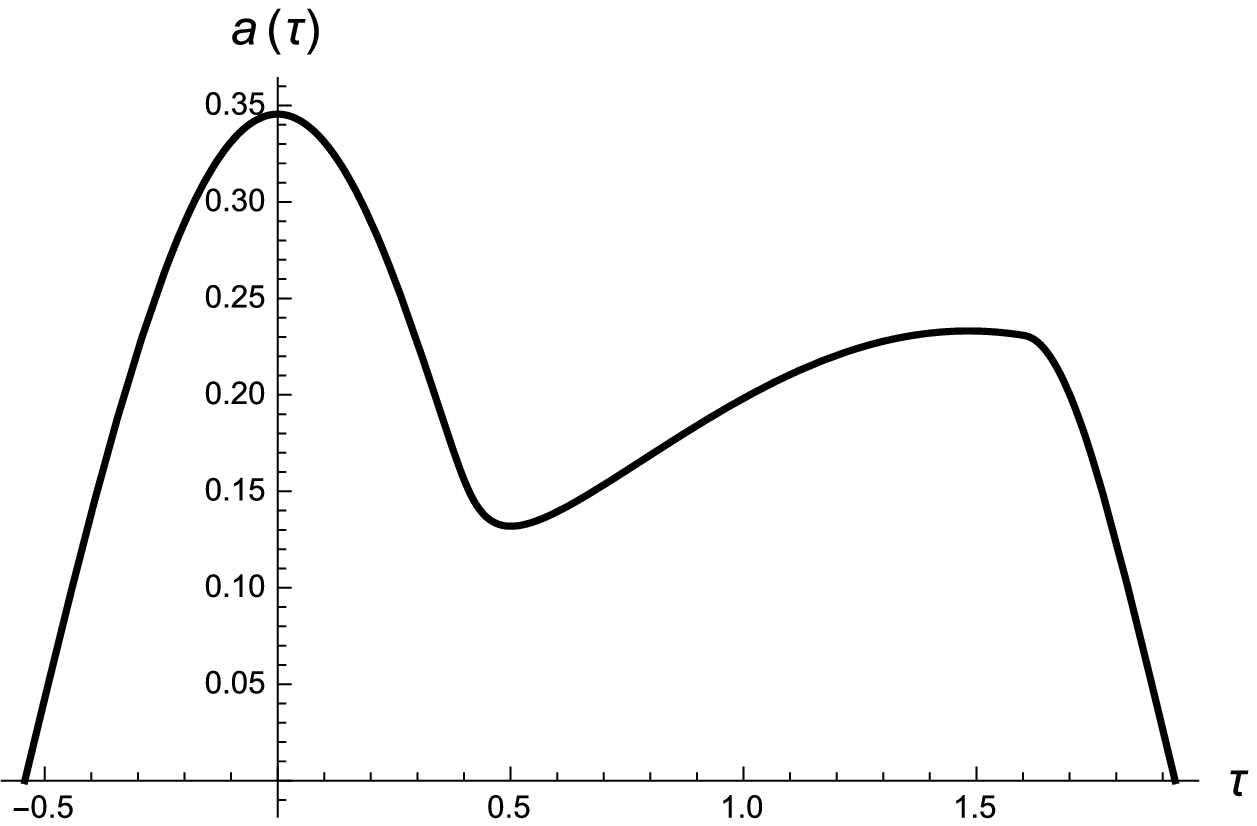}
\includegraphics[scale=0.6]{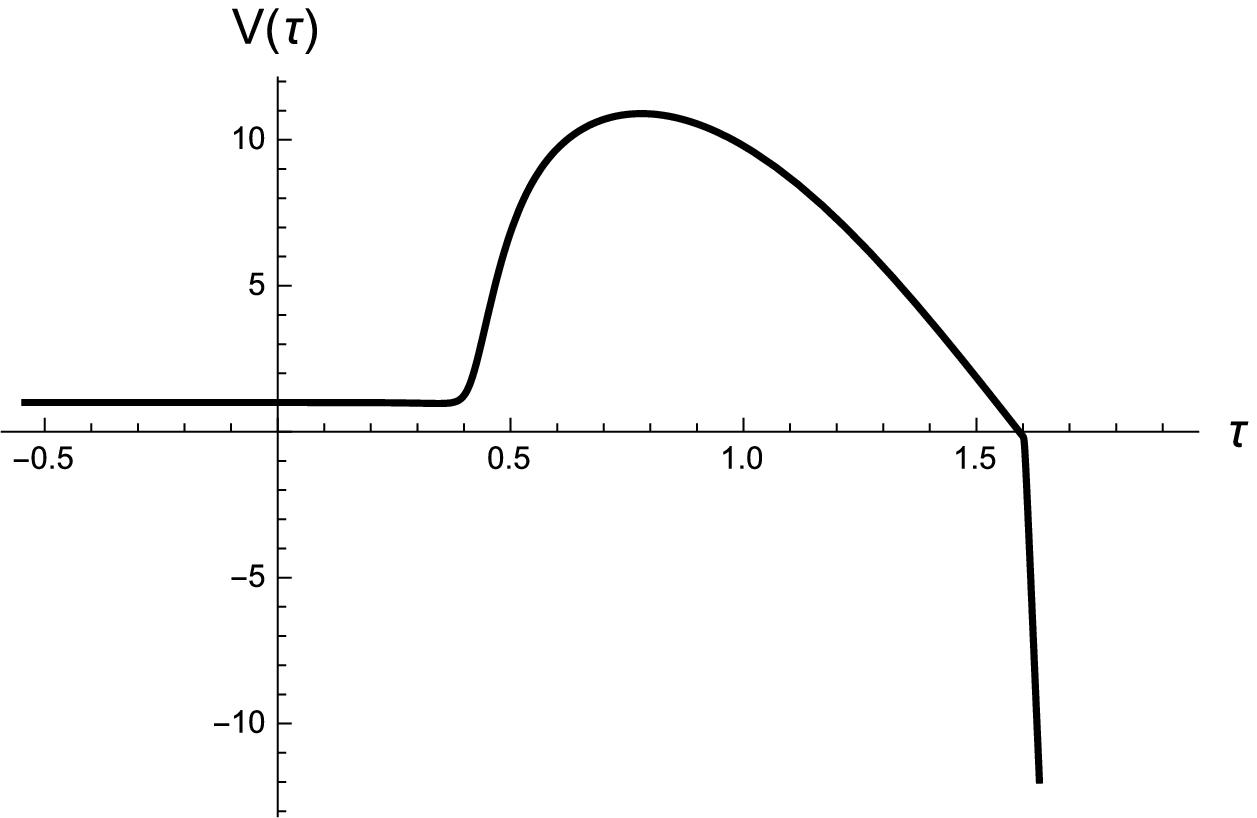}
\caption{\label{fig:demonstrate2}$a(\tau)$ (left) and $V(\tau)$ (right) for $\Delta = 1.0$.}
\includegraphics[scale=0.6]{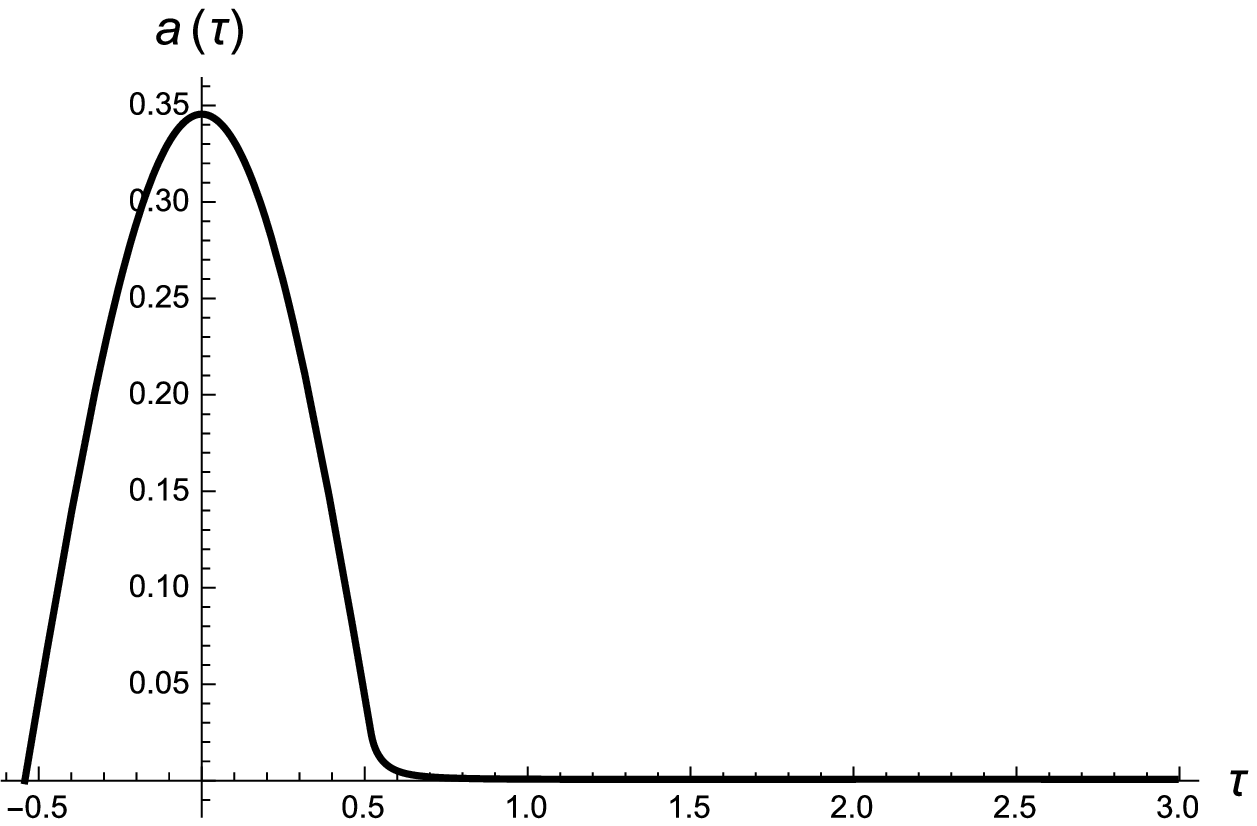}
\includegraphics[scale=0.6]{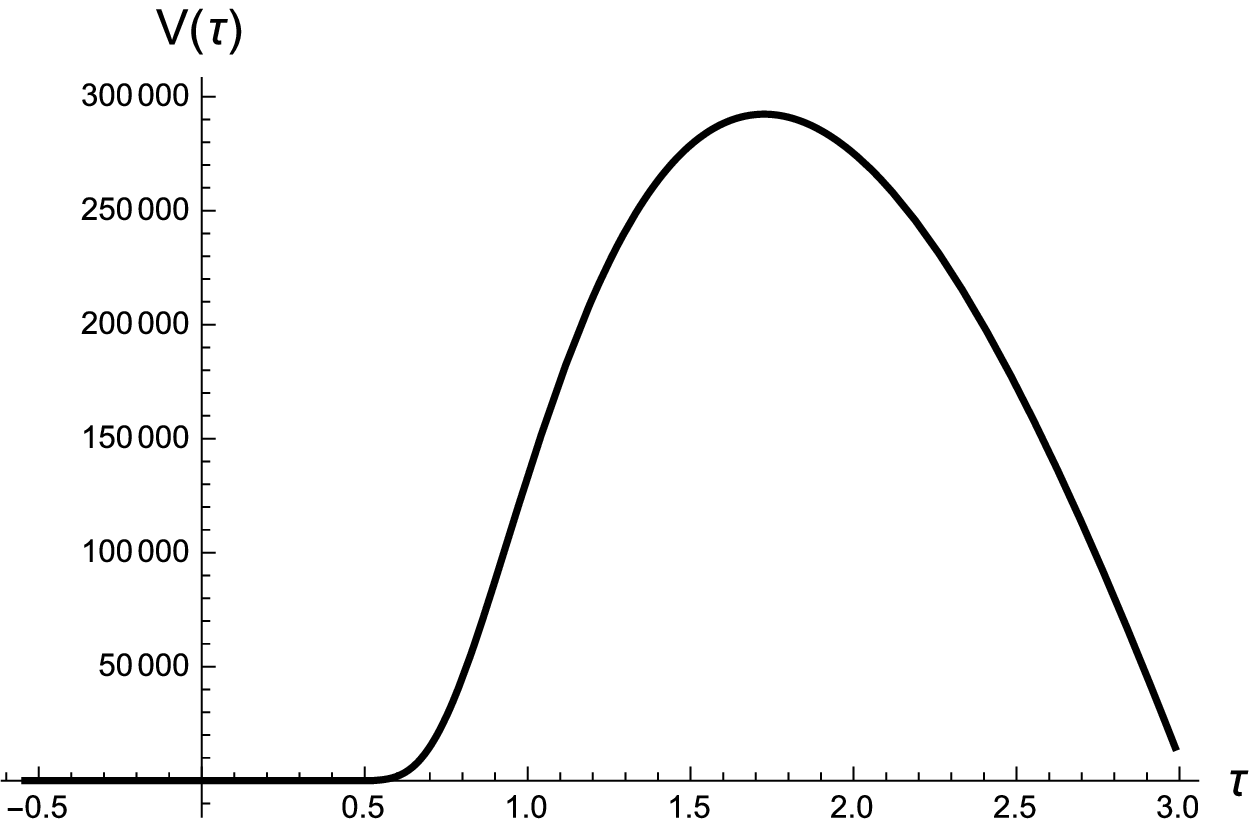}
\caption{\label{fig:demonstrate3}$a(\tau)$ (left) and $V(\tau)$ (right) for $\Delta = 1.8$.}
\end{center}
\end{figure}

\subsection{Parameter dependences}

In this subsection, we discuss more on the technical details of the solutions. Since there are many parameters that we can handle, there may be no straightforward way to finely tune the initial conditions in order to satisfy the boundary conditions for both ends. However, by changing several parameters around the solution that we have obtained, we can have some technical intuitions.

As an example, in Figs.~\ref{fig:demonstrate1}, \ref{fig:demonstrate2}, and \ref{fig:demonstrate3}, we demonstrated three examples by varying $\Delta$, where Fig.~\ref{fig:demonstrate1} is the same as Figs.~\ref{fig:atau} and \ref{fig:Vtau}. As we increase $\Delta$, the slope of $\phi$ becomes more and more gentle. Therefore, the contributions from the kinetic energy decreases and $V$ should be modified according to the back-reactions.

By comparing $V(\tau)$ of Figs.~\ref{fig:demonstrate1}, \ref{fig:demonstrate2}, and \ref{fig:demonstrate3}, we can notice that the scales of $V$ increases drastically. In Fig.~\ref{fig:demonstrate2}, $V$ crosses from positive to negative values with relatively steeper way, and hence, there appears a cusp-like point due to the sharp dynamics nearby $\tau \sim 1.7$.

One another tendency is that as $\Delta$ increases, the values of $a$ at the local minimum ($\dot{a} = 0$ and $\ddot{a} > 0$) and the second local maximum ($\dot{a} = 0$ and $\ddot{a} < 0$) become relatively lower than that of the first local maximum ($a \sim 0.35$). This means that as $\Delta$ increases, $a$ of the local minimum and the second local maximum continuously decreases and eventually there exists a limit such that $\dot{a} \rightarrow 0$ as $a \rightarrow 0$ (Fig.~\ref{fig:demonstrate3})\footnote{This behavior is quite similar to instantons in loop quantum gravity, e.g., see \cite{lqc}.}.

In this paper, we could not report all possible variations of parameters, but these examples show that there are plenty of solutions that will have interesting physical implications.

\subsection{Energy conditions}

The energy conditions arise from the Raychaudhuri equation for the expansion~\cite{Raychaudhuri:1953yv}, which is given by
\begin{eqnarray}\label{eq:RayEq1}
\frac{d\theta}{d\eta}=-\frac{1}{2}\theta^2-\sigma_{\mu\nu}\sigma^{\mu\nu}+\omega_{\mu\nu}\omega^{\mu\nu}-R_{\mu\nu}k^{\mu}k^\nu\,,
\end{eqnarray}
where $\eta$ is the affine parameter of an observer moving along a null geodesic and $R_{\mu\nu}$ is the Ricci tensor. In addition, $\theta$, $\sigma^{\mu\nu}$, and $\omega^{\mu\nu}$ are the expansion, shear, and rotation, respectively, associated to the congruence defined by the null vector field $k^\mu$.

The characteristic of attractive gravity is stated by the condition $d\theta/d\eta<0$. For the infinitesimal distortions, the Frobenius theorem \cite{ref:book1} shows that $\omega_{\mu\nu}=0$ when the congruence curves (time-like, space-like, or null) are hypersurface orthogonal. We, therefore, ignore the third term of the right-hand side of Eq.~(\ref{eq:RayEq1}). Since the first term and the second term are always negative, the attractive gravity is guaranteed if
\begin{eqnarray}\label{eq:NEC1}
R_{\mu\nu}k^\mu k^\nu\geq0\,.
\end{eqnarray}
This condition is called by the null energy condition, which is equivalent to $T_{\mu\nu}k^\mu k^\nu \geq 0$, where $T_{\mu\nu}$ is the (effective) energy-momentum tensor.

After the Wick-rotation, null vectors should be proportional to $k^\mu=(\pm1, a, 0, 0)$, and hence, by using Eq.~(\ref{eq:metric1}), the null energy condition Eq.~(\ref{eq:NEC1}) is reduced to
\begin{eqnarray}
1-\dot{a}^2+a\ddot{a}\leq0\,.
\end{eqnarray}
This is equivalent to
\begin{eqnarray}
\ddot{a}\leq \frac{1}{a}\left(\dot{a}^2-1\right).
\end{eqnarray} 
This implies that, if $\ddot{a}>0$ for $\dot{a}=0$ and $a > 0$, the null energy condition should be violated as we expected.

\begin{figure}
\begin{center}
\includegraphics[scale=0.75]{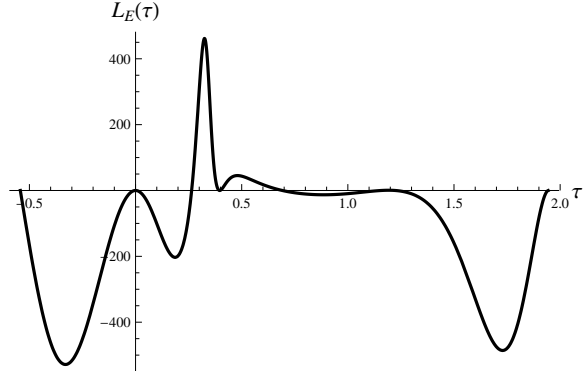}
\caption{\label{fig:lag}$L_{\mathrm{E}}(\tau)$ for the Euclidean wormhole solution.}
\end{center}
\end{figure}

\subsection{Probabilities}

Since the solution is compact, the Euclidean action should be finite and well-defined. The mini-superspace Euclidean action is \cite{Cai:2008ht}
\begin{eqnarray}
S_{\mathrm{E}} = \int L_{\mathrm{E}} d\tau,
\end{eqnarray}
where
\begin{eqnarray}
L_{\mathrm{E}} = 2\pi^{2} a^{3} \left[ \frac{\dot{\phi}^{2}}{2} + V(\phi) - \frac{3 (1 - \dot{a}^{2} - a \ddot{a})}{\kappa^{2} a^{2}} + 12 \xi(\phi) \frac{\ddot{a}}{a^{3}} \left( 1 - \dot{a}^{2} \right) \right].
\end{eqnarray}
By applying the integration by part, we obtain
\begin{eqnarray}
L_{\mathrm{E}} = 2\pi^{2} a^{3} \left[ \frac{\dot{\phi}^{2}}{2} + V(\phi) - \frac{3}{\kappa^{2}a^{2}} \left(1 + \dot{a}^{2} \right) - 12 \xi' \dot{\phi} \frac{\dot{a} (1 - \dot{a}^{2})}{a^{3}} + 24 \xi \frac{\ddot{a} \dot{a}^{2}}{a^{3}} \right].
\end{eqnarray}
Finally, by plugging Eq.~(\ref{Eq1}), we obtain
\begin{eqnarray}
L_{\mathrm{E}}^{\mathrm{on-shell}} = 4\pi^{2} \left[ a^{3} \tilde{V} - \frac{3}{\kappa^{2}} a \right],
\end{eqnarray}
where
\begin{eqnarray}
\tilde{V} = V + 12 \xi \frac{\dot{a}^{2} \ddot{a}}{a^{3}}.
\end{eqnarray}
This action is equivalent to that of Einstein gravity if $\xi = 0$.

An example of the Euclidean Lagrangian is numerically demonstrated by Fig.~\ref{fig:lag}. After the integration, the Euclidean action is negative ($-306.492$) which is quite usual for compact instantons. One may further consider the tunneling from the de Sitter space with the vacuum energy $V = V_{0}$ and $\phi = \phi_{0}$, where the solution is $a(\tau) = H_{0}^{-1} \sin H_{0} \tau$ with $H_{0}^{2} = \kappa^{2} V_{0} / 3$. For this case, the Euclidean action is
\begin{eqnarray}
S_{\mathrm{E}} = - \frac{3}{8V_{0}} \left( 1 + \frac{256 \pi^{2}}{3} \xi(\phi_{0}) V_{0} \right).
\end{eqnarray}
Hence, in our numerical setup, the action is $-316.202.$ Therefore, after subtracting to the background solution, the nucleation rate $\Gamma \propto e^{-B}$ is exponentially suppressed, i.e.,
\begin{eqnarray}
B = S_{\mathrm{E}} (\mathrm{solution}) - S_{\mathrm{E}} (\mathrm{background})
\end{eqnarray}
is positive definite.

Of course, there is a subtlety to choose the background solution. However, it is fair to say that our solution strongly supports that there can be a well-defined tunneling process from a pure de Sitter space to a time-like wormhole space within the framework of string-inspired Gauss-Bonnet-dilaton gravity.

\section{\label{sec:dis}Discussion}

In this paper, we investigated a nucleation of a Euclidean wormhole and its analytic continuation to Lorentzian signatures in Gauss-Bonnet-dilaton gravity. This model can be embedded by the type-II superstring theory. We show that there exists a Euclidean wormhole solution in this model by choosing a suitable shape of the dilaton potential. After the analytic continuation, this explains a quantum creation of a time-like wormhole \textit{ex nihilo}. This work is new since we could deal the following two topics at the same time: (1) we embedded the model in the context of the string theory\footnote{If we do not restrict to the string-inspired models, there may be further examples, e.g., Eddington-inspired-Born-Infeld gravity \cite{EiBI}.} and (2) we explained a quantum mechanical creation process of a wormhole.

Based on this solution, there may be several applications of the wormhole solution.
\begin{itemize}
\item[(1)] This solution opens a possibility that a time-like wormhole can exist in the de Sitter space, e.g., in our universe. The causal structure near the throat will give specific signals in terms of gravitational waves \cite{Volkel:2018hwb}. Also, this has an implication in the context of the open inflation scenario \cite{Yamamoto:1995sw}.
\item[(2)] In the orange colored region in Fig.~\ref{fig:concept}, there appears a hyperbolic time-like wormhole, where the other two time-like boundaries of the orange colored region correspond to the boundary of increasing areal radius. On this background, one may test some ideas of holography \cite{Maldacena:1997re}.
\end{itemize}

In addition to this, the applications to the information loss problem \cite{Hawking:1976ra,Chen:2014jwq} would be interesting. By giving matters from both sides of the wormhole throat, this wormhole can be trapped by apparent horizons. Then can the ER=EPR conjecture be still true? Since we only relied on the classical field combinations in order to have the traversable wormhole throat, we cannot rely on the quantum teleportation arguments to rescue the principle \cite{Gao:2016bin}. Also, the information loss problem is deeply related to the non-perturbative contributions of the Euclidean path-integral \cite{Hawking:2005kf,Sasaki:2014spa} and hence we need to investigate the possible effects of Euclidean wormholes \cite{ArkaniHamed:2007js}, e.g., the violation of unitarity \cite{Hawking:1987mz}. Since the Gauss-Bonnet-dilaton model can violate the effective null energy condition, there may be some hope to resolve singularities inside the black hole. We leave these topics for future investigations.

\newpage

\section*{Acknowledgment}

We would like to thank Daeho Ro for fruitful discussions. GT is pleased to appreciate Asian Pacific Center for Theoretical Physics (APCTP) for its hospitality during completion of this work. DY was supported by the Korea Ministry of Education, Science and Technology, Gyeongsangbuk-Do and Pohang City for Independent Junior Research Groups at the Asia Pacific Center for Theoretical Physics and the National Research Foundation of Korea (Grant No.: 2018R1D1A1B07049126). GT was supported by Institute for Basic Science (IBS) under the project code, IBS-R018-D1.

\end{document}